\documentclass[twocolumn,amssymb,amsmath,aps,pra,groupedaddress,nobibnotes,showpacs]{revtex4}
\usepackage{graphicx}

\begin{document}

\title{Quantum interference with distinguishable photons through indistinguishable pathways}

\author{Yoon-Ho Kim}\email{yokim@umbc.edu}
\affiliation{Center for Engineering Science Advanced Research, 
Computer Science \& Mathematics Division,
Oak Ridge National Laboratory, Oak Ridge, Tennessee 37831, U.S.A. }

\author{Warren P. Grice}
\affiliation{Center for Engineering Science Advanced Research, 
Computer Science \& Mathematics Division,
Oak Ridge National Laboratory, Oak Ridge, Tennessee 37831, U.S.A. }

\date{\today}

\begin{abstract}
We report a two-photon quantum interference experiment in which the detected individual photons have very different properties. The interference is observed even when no effort is made to mask the distinguishing features before the photons are detected. The results can only be explained in terms of indistinguishable two-photon amplitudes. 
\end{abstract}

\pacs{03.65.Bz, 42.50.Dv}

\maketitle

\section{Introduction} 
Quantum mechanical interference is observed when an event can occur by any of several alternate pathways. If an experiment is performed in which it is possible to determine which of the alternate pathways was actually taken, then the interference is lost \cite{feynman1,feynman2}. For two-photon interference, the existence of indistinguishable alternate pathways for a pair detection events leads to the interference. One of the best-known examples is the Hong-Ou-Mandel interferometer, in which two identical photons meet at a 50/50 beamsplitter and two single-photon detectors monitor the rate at which both detectors register photons (coincidence count) \cite{hong}. A coincidence count may be recorded either when both photons are reflected (r-r) or when both are transmitted (t-t). If the photons reach the beamsplitter simultaneously, then these two pathways are indistinguishable and interference is observed in the form of a photon bunching effect: the photons exit the beamsplitter together, resulting in a null in the coincidence rate \cite{kim2}. 

Although the photons reaching the beamsplitter in this example are identical, it is possible to observe interference even if the photons are distinguishable when they reach the beamsplitter. The key is to detect the photons in such a way that the distinguishable information is masked \cite{eraser}. If, for example, the input photons are orthogonally polarized, no interference is expected, since the polarizations of the detected photons would make it possible to distinguish the r-r and t-t pathways. The interference is restored, however, simply by passing the photons through polarizers prior to detection. With their pass axes oriented halfway between the polarizations of the two input photons, the polarizers mask the polarization information \cite{shih, eraser2,pittman}. The same principles may be applied to distinguishing timing information. If the photons in one input port arrive earlier than their counterparts, then the r-r and t-t pathways are distinguishable by the photon arrival times and no interference is observed. Again, the interference may be restored by masking the distinguishing timing information \cite{pla}. This may be accomplished, for example, by introducing alternate pathways into one of the exit ports \cite{pittman,delay}. The photon can reach the detector by one of two paths, leading to two different arrival times. If the delays are chosen properly, it is impossible to determine whether the photon left the beamsplitter first and took the longer path to the detector, or whether it left later and reached the detector via the shorter path. It is also possible mask the distinguishing timing information by using narrow-band spectral filters which increase inherent uncertainties associated with the differences in the photon pair arrival times at the beamsplitter \cite{kim1,filter}. 

These examples illustrate the role of indistinguishability in quantum interference. If the photons are not indistinguishable in all respects when they reach the beamsplitter, then elements must be introduced to mask the distinguishing information, in effect rendering the distinguishable photons indistinguishable. Here, we report an interference experiment in which the detected photons retain their distinguishing information. The photons approach the beamsplitter at different times, with different polarizations, and may even have different wavelengths. They propagate directly to the detectors without passing through compensating/masking elements and retain their distinguishing properties until being absorbed by the detectors. Nonetheless, interference is observed in the coincidence rate. We explain this counterintuitive result as the interference between two-photon wavepackets.

\section{Experiment}
\begin{figure}[t]
\includegraphics[width=3.1in]{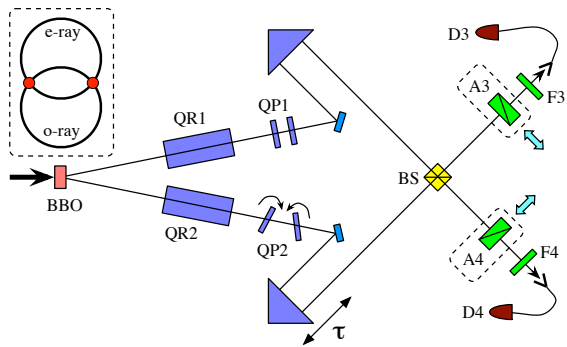}
\caption{\label{fig:setup}Outline of Experimental setup. Orthogonally polarized photon pairs are generated from the BBO crystal via the process of type-II spontaneous parametric down-conversion. Polarization analyzers A3 and A4 are removable.}
\end{figure}

An outline of the experimental setup is shown in Fig.~\ref{fig:setup}. A 3 mm thick type-II BBO crystal is pumped by a train of ultrafast pulses with central wavelength of 390 nm and pulse duration of approximately 120 fsec. The crystal is oriented so that a small fraction of the pump photons are down-converted, via the process of spontaneous parametric down-conversion (SPDC), to orthogonally polarized signal and idler photons with center wavelengths of 780 nm. The down-converted photons are emitted into two distinct cones, one with extraordinary polarization (V-polarized) and the other with ordinary polarization (H-polarized). Here, we are interested in the intersections of the two light cones, where the polarizations of the single photons cannot be defined \cite{kwiat,pulse,kim3}. These two directions, defined by a set of apertures, make an angle of $\pm$ 3 degrees with respect to the pump propagation direction, see inset of Fig.~\ref{fig:setup}. These two spatial modes are directed by mirrors to the two input ports of an ordinary non-polarizing beamsplitter. The output ports are monitored by single-photon counting detectors, D3 and D4. Broadband (20 nm at FWHM) spectral filters, F3 and F4, preceding the detectors help to reduce background noise. Polarization analyzers A3 and A4 are removable so that the interference effect can be studied both with and without the polarizers in place.

A set of quartz rods and quartz plates are inserted in each arm of the interferometer: QR1 and QR2 are 20 mm long quartz rods and QP1 and QP2 are 600 $\mu$m thick quartz plates. With their optic axes oriented vertically, these birefringent elements introduce a group delay of roughly 668 fsec between the V- and H-polarized photons in addition to the delay accumulated in the SPDC crystal. By tilting the quartz plates QP2 about their optic axes, it is possible to introduce an additional fine delay between the orthogonally polarized photons. The delay between the two arms of the interferometer is controlled by a trombone prism attached to a computer-controlled DC motor. The count rates of the two detectors, as well as the rate of coincidences, were recorded as a function of the delay between the two arms of the interferometer. The effective coincidence window used in this experiment was about 3 nsec, which is smaller than the pump pulse repetition period ($\approx$ 13 nsec). 

\begin{figure}[t]
\includegraphics[width=3.2in]{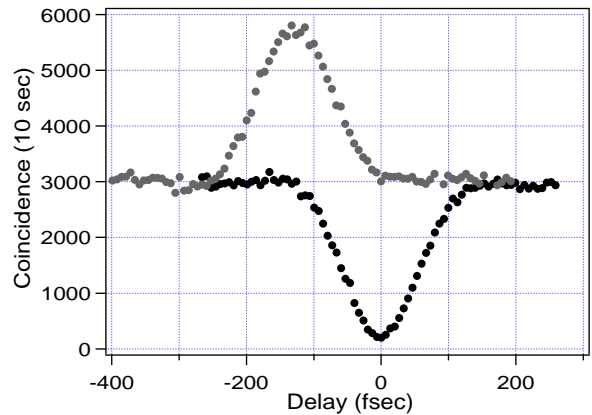}
\caption{\label{fig:data1}Experimental data showing two-photon quantum interference. Polarization analyzers were not used for this measurement. The peak-dip visibility is about 93\%.}
\end{figure}

Quantum interference is observed as the delay between the two arms is adjusted. This can be seen as the ÒpeakÓ and ÒdipÓ shown in Fig.~\ref{fig:data1}. The two different data sets correspond to two different ÒphaseÓ settings, i.e., two different orientations of the quartz plates QP2. Tilting the quartz plates introduces a sub-wavelength delay between the orthogonally polarized modes in the lower arm. The peak-dip phase is associated with this additional birefringent delay. Tilting the quartz plates increases not only the relative delay, but also the total path length for the two different polarizations. This is reflected as an offset between the peak and dip.

We note here that the photon pairs approach the beamsplitter with very different properties: they are orthogonally polarized and approach at different times (one photon always reaches the beamsplitter 668 fsec earlier than the other). This time delay is larger than both the 120 fsec pump pulse duration and the 100 fsec single-photon wave-packet defined by the 20 nm (at FWHM) filter. It is somewhat surprising, then, that high visibility interference is observed even though the photons' properties are not altered before the detection process, i.e., no elements are introduced after the beamsplitter to mask the distinguishing information. (For instance, $45^\circ$ oriented polarizers and a group delay scheme could be used to `erase' the polarization and temporal distinguishing information, respectively.)

\begin{figure}[t]
\includegraphics[width=3.2in]{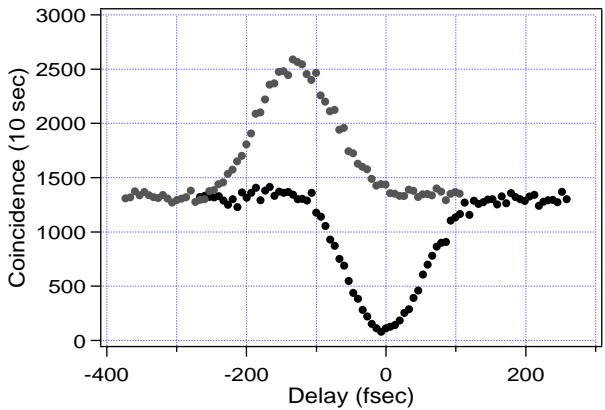}
\caption{\label{fig:data2}Experimental data with polarizers oriented at  A3 $=90^\circ$ and A4 $=0^\circ$. The visibility of quantum interference remains the same but the overall count rate has been reduced.}
\end{figure}

To emphasize the fact that the quantum interference is indeed achieved with distinguishable photons, the experiment was repeated with polarizers A3 and A4 placed in front of the detectors. One of the polarizers was aligned to pass horizontally polarized light (A3 = $90^\circ$), while the other was aligned to pass vertically polarized light (A4 = $0^\circ$). The only photons that may be detected in such a set-up are emitted and detected with orthogonal polarizations and are orthogonally polarized at every point in the interferometer. Nevertheless, high visibility quantum interference is observed, as shown in Fig.~\ref{fig:data2}. Although the overall count rate is lowered, the peak and dip features are still evident with no change in the visibility. 

In fact, the peak-dip quantum interference is completely independent of the polarizer angles. With QP2 fixed for the coincidence peak (dip), the measurement was repeated for polarizer angles A3 $= 45^\circ$ and A4 $=-45^\circ$ (orthogonal polarizers) and A3 $= 45^\circ$ and A4 $=45^\circ$ (parallel polarizers). These polarizer settings will typically change the coincidence peak (dip) into a coincidence dip (peak) \cite{shih,eraser2,pittman}. In this experiment, however, the peak structure remained, with no change in the visibility: only the overall count rates are reduced. Figure \ref{fig:data3} shows the data for this set of measurements. 

\section{Theory} 
The experimental results make it clear that it is possible to observe high-visibility quantum interference, even when the detected photons have very different properties. The interference effect can be understood more clearly through a calculation of the coincidence rate at the two detectors. As described above, the photons leave the SPDC crystal in the paths corresponding to the overlap of the o- and e-polarized cones. The two-photon state may be written in this case as
\begin{equation}
|\Psi\rangle =\frac{1}{\sqrt{2}}\left\{ {|\psi _{HV}\rangle +e^{i\phi }|\psi _{VH}\rangle } \right\} \label{eq:state}
\end{equation}
where $|\psi_{HV}\rangle$ and $|\psi_{VH}\rangle$ represent the two ways in which a photon pair may be emitted into these directions and where $\phi$ represents the phase between the two terms. Because the crystal is aligned for type-II SPDC and because it is pumped by a train of short pulses, a multi-mode treatment is required in order to reveal the subtle spectral and temporal characteristics of the two-photon state. 

\begin{figure}[t]
\includegraphics[width=3.2in]{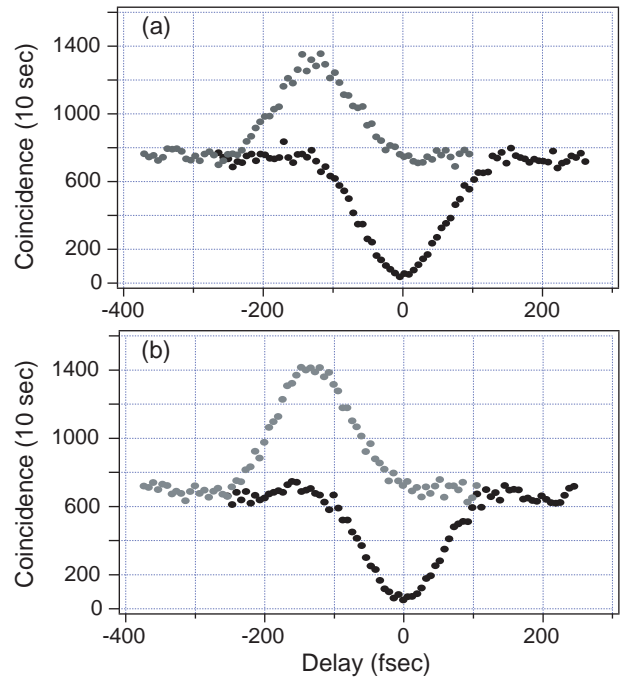}
\caption{\label{fig:data3} Experimental data with (a) polarizers A3 $= 45^\circ$ and A4 $=45^\circ$, (b) polarizers A3 $=45^\circ$ and A4 $=-45^\circ$. The overall count rate has been reduced but the visibility remains the same.}
\end{figure}

Although the state may be described in either the spectral or temporal domain, the analysis in this case is somewhat simpler in the temporal domain. Accordingly, the states $|\psi_{HV}\rangle$ and $|\psi_{VH}\rangle$ are
\begin{eqnarray}
|\psi_{HV}\rangle =\iint dt_o \, dt_e \, \mathcal{G}_{HV}(t_o,t_e)\hat a_{H1}^\dagger(t_o) \hat a_{V2}^\dagger (t_e)|0\rangle, \nonumber\\
|\psi_{VH}\rangle =\iint dt_o \, dt_e \, \mathcal{G}_{VH}(t_o,t_e)\hat a_{V1}^\dagger (t_e) \hat a_{H2}^\dagger (t_o)|0\rangle, \nonumber
\end{eqnarray}
where the two photon amplitude is given by 
\begin{eqnarray}
&& \mathcal{G}_{HV}(t_o,t_e)= \mathcal{G}_{VH}(t_o,t_e)\nonumber\\ &&= N e^{-i\bar \omega (t_o+t_e)}\xi (t_o,t_e)\Pi \left[ {t_e-t_o;0,L(k'_o-k'_e)} \right].\nonumber
\end{eqnarray}
Here, $N$ is a normalization constant, ${\bar \omega }$ is the mean photon frequency, $L$ is the crystal length, and the function $\Pi \left[ {t;t_1,t_2} \right]$ is given by
\begin{equation}
\Pi[t;t_1,t_2]=\left\{ \begin{array}{l}
						1\ \ \ \ \ \textrm{for} \ \ \ \ t_1<t<t_2\\	
						0\ \ \ \ \ \ \ \ \ \ \ \ \textrm{otherwise}.
					   \end{array}\right.
\end{equation}

The function $\xi \left( {t_o,t_e} \right)$ is related to the pump field and the birefringent properties of the crystal. For a pump field proportional to $\exp \{-[ (\omega -2\bar \omega)/\sigma]^2\}$, $\xi \left( {t_o,t_e} \right)$ is  
$$
\xi(t_o,t_e)=\exp \left\{ {-\frac{\sigma ^2}{4} \left[ {\left( \frac{k'_p-k'_e}{k'_o-k'_e} \right)t_o-\left( \frac{k'_p-k'_o}{k'_o-k'_e} \right)t_e} \right]^2} \right\}
$$
where $k'_p=\left. \frac{dk_p}{d\omega} \right|_{2\bar \omega}$ 
and $k'_{o(e)}=\left. \frac{dk_{o(e)}}{d\omega} \right|_{\bar \omega} $
are the inverse group velocities of the pump ($p$), o-polarized ($o$), and e-polarized ($e$) photons, respectively. 

Expressed in this way, it is easy to see that the two-photon probability amplitude describes a pair of photons whose emission times are determined primarily by the temporal shape of the pump pulse, modified somewhat by the group velocity differences inside the crystal. The rectangle function $\Pi \left[ {t;t_1,t_2} \right]$ sets upper and lower bounds for the difference in emission times and reflects the fact that the o- and e-polarized photons will separate temporally as they propagate through the crystal. With a lower bound of zero (for a pair created at the exit face) and an upper bound of ${L\left( {k'_o-k'_e} \right)}$ (for a pair created at the entrance face), it is evident that these expressions describe the two-photon states as they exit the crystal, i.e., before the photons pass through any birefringent elements.

After exiting the crystal, the photons propagate along paths 1 and 2, experiencing nearly identical birefringent delays (due to the quartz rods and quartz plates). They are then brought together at a beamsplitter, though the path lengths may be slightly different. To within a constant overall phase, the two-photon state at the input of the beamsplitter is simply the state given in Eq. (1) with the temporal arguments shifted as follows:
\begin{eqnarray}
\mathcal{G}_{HV}(t_o,t_e) \to \mathcal{G}_{HV}(t_o,t_e-\tau _2-\tau) \nonumber \\
\mathcal{G}_{VH}(t_o,t_e) \to \mathcal{G}_{VH}(t_o-\tau ,t_e-\tau _1) \nonumber
\end{eqnarray}
Here, $c\tau $ is the difference in free-space path lengths between the upper and lower paths. The delays $\tau _1$ and $\tau _2$ represent the delays of the e-polarized photons with respect to the o-polarized photons in paths 1 and 2, respectively.

With no polarizers in place, the coincidence rate at detectors D3 and D4 is given by 
\begin{equation}
R=\iint dt \, dt' \, \{ P_{HV}(t,t') + P_{VH}(t,t') \}, \label{eq:coinc1}
\end{equation}
where ${P_{HV}\left( {t,t'} \right)}$ [${P_{VH}\left( {t,t'} \right)}$] is the probability that a horizontally [vertically] polarized photon is detected at D3 at time $t$ and that a vertically [horizontally] polarized photon is detected at D4 at time ${t'}$. The expression should in general also include the terms ${P_{HH}\left( {t,t'} \right)}$ and ${P_{VV}\left( {t,t'} \right)}$, but since the photons emitted in type-II SPDC are orthogonally polarized, these are zero for the experiment described here.  

The two-time detection probabilities are given by 
$$
P_{ij}(t,t')=\left| \hat a_{i3}(t) \hat a_{j4}(t') |\Psi\rangle\right|^2,
$$
where $i$ and $j$ are the polarization labels which can be $H$ or $V$ depending on the polarization state of the photon in a given spatial mode. To within a constant phase factor, the annihilation operators at the detectors are related to the input operators by
\begin{eqnarray}
\hat a_{i3}(t)= \frac{1}{\sqrt{2}} [\hat a_{i2}(t)+i\hat a_{i1}(t) ], \nonumber \\
\hat a_{i4}(t)= \frac{1}{\sqrt{2}} [\hat a_{i1}(t)+i\hat a_{i2}(t) ]. \nonumber
\end{eqnarray}

Using the expressions given above, the coincidence count rate $R$ given in eq.~(\ref{eq:coinc1}) is found to be 
\begin{widetext}
\begin{eqnarray}
R &=& R_0 \left\{ 1-\cos[\phi -\bar \omega (\tau_2-\tau_1)] \left(1- \frac{|2\tau +\tau_2-\tau _1|}{L (k'_o-k'_e)}\right) \right. \nonumber\\ &&\left. \times \exp \left[ -\frac{\sigma ^2}{8} \left[ \left(\frac{2k'_p-k'_o-k'_e}{k'_o-k'_e}\right) \tau + \left(\frac{k'_p-k'_o}{k'_o-k'_e} \right) (\tau _2-\tau _1) \right]^2 \right] \Pi[2\tau +\tau _2-\tau _1; -L (k'_o-k'_e),L(k'_o-k'_e)] \right\},
\end{eqnarray}
\end{widetext}
where $R_0$ is a constant. While this expression is a bit cumbersome, it simplifies greatly for $\tau =0$ (equal path lengths) and $\tau _1-\tau _2\approx 0$ (nearly identical birefringent delays), in which case the expression becomes
\begin{equation}
R\approx R_0 \{ 1-\cos[\phi -\bar \omega ( \tau _2-\tau _1 )] \}. \label{eq:final}
\end{equation}
In this form, it is clear that the coincidence count rate can be made to vanish for the appropriate birefringent delays and, likewise, that it can be raised to twice the background rate with a small phase shift. [Recall that $\phi$ is the unknown fixed phase term from eq.~(\ref{eq:state}).] It is these two settings that were used to generate the data shown in Figs.~\ref{fig:data1} $\sim$ \ref{fig:data3}. As described above, the phase shift in $\tau _1-\tau _2$ was introduced in our experiment by tilting QP2. This type of adjustment introduces slightly different path length increases for the two polarizations, but also increases the overall length of the lower arm (of the interferometer) so that the trombone prism must be adjusted to compensate. This is the origin of the offset between the coincidence peak and the coincidence dip observed in the figures.

\section{Discussions}

A couple of results from the above analysis invite further comment. First, the interference features are independent of the presence of polarizers or, if polarizers are in place, of their orientations. This is in contrast to previous experiments, where the relative orientations of the polarizers determine the phase between the interfering terms and whether the interference is constructive or destructive. The inclusion of polarizers in the above analysis changes the overall rate of coincidence detection, but the peak/dip features persist. Thus, photon polarization has absolutely no bearing on the interference effect and there is no need to mask differences between the photonsÕ polarizations.

Another result worth noting is that the photons may reach the beamsplitter at very different times \cite{pla}. Recall that the photons are emitted with orthogonal polarizations and, therefore, travel at different speeds through the birefringent elements. By adjusting the birefringent delays, it is possible to change the arrival times for the orthogonally polarized photons. For maximum visibility, $\tau_1$ and $\tau_2$ must be nearly identical, but the constraint is on their difference, not on the individual delays. Thus, interference can be observed for arbitrary amounts of birefringent delay, as long as the delays in the two arms are nearly identical.
These results stand in contrast to previous works involving similar configurations where interference could be observed only if there were no distinguishing timing or polarization information, or if such information were ÒerasedÓ before detection \cite{shih,eraser2,pittman,pulse,delay,filter}. Here, such information seems not to matter. 

\begin{figure}[tb]
\includegraphics[width=3in]{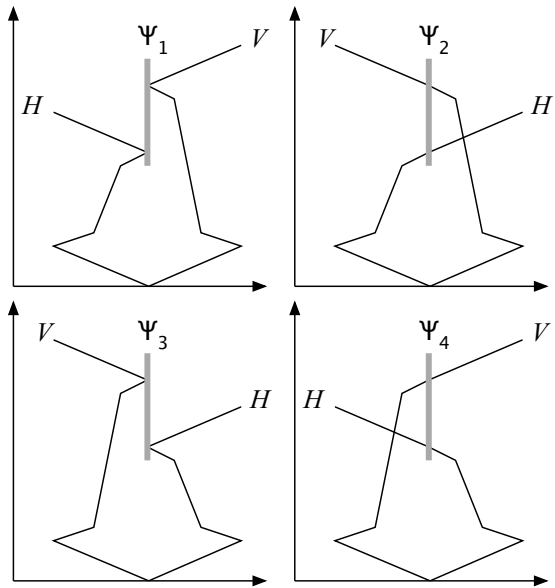}
\caption{\label{fig:amplitudes}Quantum mechanical two-photon Feynman alternatives. $\Psi_{1}$ and$\Psi_{4}$ are indistinguishable and  $\Psi_{2}$ and$\Psi_{3}$ are indistinguishable. As a result, quantum interference occur pair-wise and interference peak-dip can be observed without needing to ``erase" the actively distinguishing information. Vertical gray bar represents the beamsplitter.}
\end{figure}

The difference between those experiments and the one reported here is the input state. Whereas the previous experiments involved single two-photon states, the input state for our experiment is the superposition state of Eq.~(\ref{eq:state}). This seemingly subtle change to the input state is critical to the observed interference effect, for this superposition now provides two pathways for a given detection outcome. Suppose, for example, that detector D3 registers a horizontally polarized photon, while D4 registers a vertically polarized photon. There are two ways that this may happen: the photons may be emitted as  $|H\rangle_1|V\rangle_2$ and be reflected at the beamsplitter; or they may be emitted as $|V\rangle_1|H\rangle_2$ and be transmitted. As long as the interferometer is properly adjusted, these two pathways will be indistinguishable, even though the photons themselves may have very different properties (see $\Psi_1$ and $\Psi_4$ in Fig.~\ref{fig:amplitudes}).

The relationships between the various detection pathways are shown in the two-photon Feynman diagrams of Fig.~\ref{fig:amplitudes}. The four diagrams leading to coincident detection correspond to two possible outcomes (both photons reflected, r-r, or transmitted, t-t, at the beamsplitter) for each of the two possible emission states ($|H\rangle_1|V\rangle_2$ or $|V\rangle_1|H\rangle_2$). The four cases are pairwise indistinguishable: $\Psi_{1}$ and $\Psi_{4}$ correspond to H at D3 and V at D4, while  $\Psi_{2}$ and $\Psi_{3}$ correspond to V at D3 and H at D4. Note, also, that the horizontally polarized photon is always detected first, a consequence of the birefringent delays. As long as the delays are identical, the amplitudes remain pairwise indistinguishable.

Returning to the comparison with previous two-photon interference experiments, it is clear that indistinguishability still plays an important role, even when the photons, themselves are distinguishable. While the detected photons may be different, it is critical that the r-r and t-t pathways be indistinguishable. Typically, this condition could be satisfied only if the photons were identical as they entered the beamsplitter or if they were detected in such a way that they appeared to be identical. Here, the indistinguishability is found in the ambiguity of the input to the beamsplitter. For example, a photon with a particular set of properties may be detected at a given detector, with a second photon having a distinct set of properties at the other detector. The r-r and t-t pathways remain indistinguishable, however, because there are two ways that these two photons may be emitted into the set-up. 

It is the superposition of the two two-photon states before they reach the beamsplitter, therefore, that is essential to the interference effect. Indeed, this type of arrangement was originally proposed by Braunstein and Mann as a means of distinguishing one of the four polarization Bell states \cite{braunstein,bellmeasure}. It is interesting to note that the interference effect persists even when $|\Psi\rangle$ exhibits little or no polarization entanglement. It is well documented that a pure polarization-entangled state may be generated only when the two photons in a particular path are distinguishable \textit{only} by their polarizations. Differences in spectral properties and/or time of arrival tend to blur the entanglement \cite{pulse,kim3,kim4}. In this experiment, however, the interference effect is observed even when the photons may be distinguished by their arrival times. Since the photon pairs are generated in an ultrafast-pulse-pumped type-II crystal, they also posses distinguishing spectral information \cite{pulse,kim3,kim4,grice}. Even so, the visibility is much higher, 100\% in principle, than would be expected if this spectral information were to play a role. This suggests that it is not even necessary that the photons have the same center wavelengths.

In conclusion, we have demonstrated a two-photon quantum interference effect in which the detected individual photons have very different properties. Unlike other interference experiments, in which the photons themselves must be indistinguishable upon detection, the individual photons here have different polarization states, different arrival times, and different spectra. Indistinguishability still plays a critical role for quantum interference to occur, however, and comes in the form of the input state: for each detection event, there are essentially two possible sources (or quantum mechanical amplitudes) for the photon pair.

\section*{Acknowledgements}

This research was supported in part by the National Security Agency, the Advanced Research \& Development Activity, and the LDRD program of The Oak Ridge National Laboratory, managed for the U.S. DOE by UT-Battelle, LLC, under contract No.~DE-AC05-00OR22725.

\end{document}